\documentclass{aa}  
\usepackage{color}
\usepackage[dvipsnames]{xcolor}
\usepackage{graphicx}	
\usepackage{xspace}	    

\usepackage{txfonts}
\usepackage[colorlinks=true, citecolor=blue]{hyperref}
\begin{document} 

\newcommand{\vdag}{(v)^\dagger}
\newcommand{\myemail}{skywalker@galaxy.far.far.away}

\newcommand{\simgt}{\,\rlap{\lower 3.5 pt \hbox{$\mathchar \sim$}} \raise 1pt \hbox {$>$}\,}
\newcommand{\simlt}{\,\rlap{\lower 3.5 pt \hbox{$\mathchar \sim$}} \raise 1pt \hbox {$<$}\,}
\newcommand{\dd}{\mathrm{d}}

\newcommand{\BE}{\begin{equation}}
\newcommand{\EE}{\end{equation}}
\newcommand{\BEA}{\begin{eqnarray}}
\newcommand{\EEA}{\end{eqnarray}}

\newcommand{\Ob}{\Omega_\textrm{b}}
\newcommand{\Om}{\Omega_\textrm{m}}
\newcommand{\OL}{\Omega_\Lambda}
\newcommand{\rhoc}{\rho_\textrm{c}}

\newcommand{\CHII}{C_\textrm{\ion{H}{2}}}
\newcommand{\DV}{\ifmmode{\Delta v}\else $\Delta v$\xspace\fi}

\newcommand{\Tigm}{{\mathcal{T}_\textsc{igm}}}

\newcommand{\HI}{\ifmmode{\textsc{hi}}\else H\textsc{i}\fi\xspace}
\newcommand{\HII}{\ifmmode{\textsc{hii}}\else H\textsc{ii}\fi\xspace}
\newcommand{\OII}{[O\textsc{ii}]}
\newcommand{\OIII}{O\textsc{iii}}
\newcommand{\CIV}{C\textsc{iv}}
\newcommand{\HeII}{He\textsc{ii}}
\newcommand{\Ha}{\ifmmode{\mathrm{H}\alpha}\else H$\alpha$\fi\xspace}
\newcommand{\OiiiHb}{\ifmmode{\mathrm{[OIII]+H}\beta}\else [OIII]+H$\beta$\fi\xspace}
\newcommand{\Msun}{\ifmmode{M_\odot}\else $M_\odot$\xspace\fi}
\newcommand{\MUV}{\ifmmode{M_\textsc{uv}}\else $M_\textsc{uv}$\xspace\fi}
\newcommand{\fesc}{\ifmmode{f_\mathrm{esc}}\else $f_\mathrm{esc}$\xspace\fi}
\newcommand{\lya}{\ifmmode{\mathrm{Ly}\alpha}\else Ly$\alpha$\xspace\fi}
\newcommand{\vrot}{v_\textrm{rot}}

\newcommand{\nh}[1][]{\ifmmode{\overline{n}_\textsc{h}^{#1}}\else $\overline{n}_\textsc{h}$\xspace\fi}

\newcommand{\xHI}{\ifmmode{x_\HI}\else $x_\HI$\xspace\fi}
\newcommand{\xHImean}{\ifmmode{\overline{x}_\HI}\else $\overline{x}_\HI$\xspace\fi}
\newcommand{\xHIImean}{\ifmmode{\overline{x}_\HII}\else $\overline{x}_\HII$\xspace\fi}
\newcommand{\trec}{\ifmmode{t_\textrm{rec}}\else $t_\textrm{rec}$\xspace\fi}
\newcommand{\clump}[1][]{\ifmmode{C_\HII^{#1}}\else $C_\HII$\xspace\fi}
\newcommand{\xiion}{\ifmmode{\xi_\mathrm{ion}}\else $\xi_\mathrm{ion}$\xspace\fi}

\newcommand{\Nion}{\ifmmode{\dot{N}_{\mathrm{ion}}}\else $\dot{N}_\mathrm{ion}$\xspace\fi}
\newcommand{\Rion}[1][]{\ifmmode{R_\mathrm{ion}^{#1}} \else $R_\mathrm{ion}$\xspace\fi}

\newcommand{\Tb}{\ifmmode{T_{21}}\else $T_{21}$\xspace\fi}
\newcommand{\aesc}{\ifmmode{\alpha_\mathrm{esc}}\else $\alpha_\mathrm{esc}$\xspace\fi}
\newcommand{\fescII}{\ifmmode{f_\mathrm{esc,10}^\textsc{ii}}\else $f_\mathrm{esc,10}^\textsc{ii}$\xspace\fi}
\newcommand{\fescIII}{\ifmmode{f_\mathrm{esc,7}^\textsc{ii}}\else $f_\mathrm{esc,7}^\textsc{iii}$\xspace\fi}
\newcommand{\astarII}{\ifmmode{\alpha_\star^\textsc{ii}}\else $\alpha_\star^\textsc{ii}$\xspace\fi}
\newcommand{\astarIII}{\ifmmode{\alpha_\star^\textsc{iii}}\else $\alpha_\star^\textsc{iii}$\xspace\fi}
\newcommand{\fstarII}{\ifmmode{f_{\star,10}^\textsc{ii}}\else $f_{\star,10}^\textsc{ii}$\xspace\fi}
\newcommand{\fstarIII}{\ifmmode{f_{\star,7}^\textsc{iii}}\else $f_{\star,7}^\textsc{iii}$\xspace\fi}
\newcommand{\tstar}{\ifmmode{t_\star}\else $t_\star$\xspace\fi}
\newcommand{\Mturn}{\ifmmode{M_\mathrm{turn}}\else $M_\mathrm{turn}$\xspace\fi}
\newcommand{\LX}{\ifmmode{L_X/{\dot{M}_\star}}\else $L_X/{\dot{M}_\star}$\xspace\fi}
\newcommand{\nuX}{\ifmmode{E_0}\else $E_0$\xspace\fi}
\newcommand{\AVCB}{\ifmmode{A_\mathrm{VCB}}\else $A_\mathrm{VCB}$\xspace\fi}
\newcommand{\ALW}{\ifmmode{A_\mathrm{LW}}\else $A_\mathrm{LW}$\xspace\fi}
\newcommand{\Mpcinv}{\ifmmode{\,\mathrm{Mpc}^{-1}}\else \,Mpc$^{-1}$\xspace\fi} 

\newcommand{\kp}{\ifmmode{k_\textrm{peak}}\else $k_\textrm{peak}$\xspace\fi}
\newcommand{\hp}{\ifmmode{h_\textrm{peak}}\else $h_\textrm{peak}$\xspace\fi}
\newcommand{\hMpc}{\ifmmode{\,h^{-1}\textrm{Mpc}}\else \,$h^{-1}$Mpc\xspace\fi}

\newcommand{\fdens}{\,erg s$^{-1}$ cm$^{-2}$\xspace}
\newcommand{\kms}{\,\ifmmode{\mathrm{km}\,\mathrm{s}^{-1}}\else km\,s${}^{-1}$\fi\xspace}
\newcommand{\cm}{\,\ifmmode{\mathrm{cm}}\else cm\fi\xspace}

\newcommand{\HST}{\textit{HST}}
\newcommand{\JWST}{\textit{JWST}}
\newcommand{\WFIRST}{\textit{WFIRST}}

\newcommand{\NB}[1]{\textbf{\color{red} #1}}
\newcommand{\tnm}[1]{$^\textrm{#1}$}

   \title{The production of ionizing photons in UV-faint $z\sim3-7$ galaxies}

   \author{
        Gonzalo Prieto-Lyon\fnmsep\thanks{E-mail: gonzalo.prieto@nbi.ku.dk}\inst{1,2}
        \and Victoria Strait\inst{1,2}
        \and Charlotte A. Mason\inst{1,2}
        \and Gabriel Brammer\inst{1,2}
        \and Gabriel B.~Caminha\inst{3,4}
        \and Amata Mercurio\inst{5,6}
        \and Ana Acebron\inst{7,8}
        \and Pietro Bergamini\inst{7,9}
        \and Claudio Grillo\inst{7,8}
        \and Piero Rosati\inst{9,10}
        \and Eros Vanzella\inst{10}
        \and Marco~Castellano\inst{11}
        \and Emiliano Merlin\inst{11}
        \and Diego Paris\inst{11}
        \and Kristan~Boyett\inst{12,13}
        \and Antonello Calabr\`o\inst{11}
        \and Takahiro Morishita\inst{14}
        \and Sara Mascia\inst{11}
        \and Laura Pentericci\inst{11}
        \and Guido Roberts-Borsani\inst{15}
        \and Namrata Roy\inst{16}
        \and Tommaso Treu\inst{15}
        \and Benedetta Vulcani\inst{17}
        }
        
   \institute{Cosmic Dawn Center (DAWN)
   \and Niels Bohr Institute, University of Copenhagen, Jagtvej 128, DK-2200 Copenhagen N, Denmark
    \and Technische Universit\"at M\"unchen, Physik-Department, James-Franck Str. 1, 85748 Garching, Germany
    \and Max-Planck-Institut für Astrophysik, Karl-Schwarzschild-Str. 1, 85748 Garching, Germany
    \and Dipartimento di Fisica “E.R. Caianiello”, Universit\`a Degli Studi di Salerno, Via Giovanni Paolo II, I–84084 Fisciano (SA), Italy
    \and INAF -- Osservatorio Astronomico di Capodimonte, Via Moiariello 16, I-80131 Napoli, Italy
    \and Dipartimento di Fisica, Universit\`a  degli Studi di Milano, via Celoria 16, I-20133 Milano, Italy
    \and INAF - IASF Milano, via A. Corti 12, I-20133 Milano, Italy
    \and INAF -- OAS, Osservatorio di Astrofisica e Scienza dello Spazio di Bologna, via Gobetti 93/3, I-40129 Bologna, Italy
    \and Dipartimento di Fisica e Scienze della Terra, Universit\`a di Ferrara, Via Saragat 1, I-44122 Ferrara, Italy
    \and INAF Osservatorio Astronomico di Roma, Via Frascati 33, 00078 Monteporzio Catone, Rome, Italy
    \and School of Physics, University of Melbourne, Parkville 3010, VIC, Australia
    \and ARC Centre of Excellence for All Sky Astrophysics in 3 Dimensions (ASTRO 3D), Australia
   \and IPAC, California Institute of Technology, MC 314-6, 1200 E. California Boulevard, Pasadena, CA 91125, USA
   \and Department of Physics and Astronomy, University of California, Los Angeles, 430 Portola Plaza, Los Angeles, CA 90095, USA
   \and Center for Astrophysical Sciences, Department of Physics \& Astronomy, Johns Hopkins University, Baltimore, MD 21218, USA
   \and INAF Osservatorio Astronomico di Padova, vicolo dell'Osservatorio 5, 35122 Padova, Italy
   }

   \date{Submitted November 22, 2022}

  \abstract
   {}
   {The demographics of the production and escape of ionizing photons from UV-faint early galaxies is a key unknown in discovering the primary drivers of reionization. With the advent of JWST it is finally possible to observe the rest-frame optical nebular emission from individual sub-$L^*$ $z>3$ galaxies to measure the production rate of ionizing photons, $\xi_\mathrm{ion}$.}
   {Here we study a sample of 370 $z\sim3-7$ galaxies spanning $-23 < M_\mathrm{UV} < -15.5$ (median $M_\mathrm{UV}\approx -18$) with deep multi-band HST and JWST/NIRCam photometry covering the rest-UV to optical from the GLASS and UNCOVER JWST surveys. Our sample includes 102 galaxies with Lyman-alpha emission detected in MUSE spectroscopy. We use H$\alpha$ fluxes inferred from NIRCam photometry to estimate the production rate of ionizing photons which do not escape these galaxies $\xi_\mathrm{ion}(1-f_\mathrm{esc})$.}
   {We find median $\log_{10}\xi_\mathrm{ion}(1-f_\mathrm{esc})=25.33\pm 0.47$, with a broad intrinsic scatter 0.42\,dex, implying a broad range of galaxy properties and ages in our UV-faint sample. Galaxies detected with Lyman-alpha have $\sim0.1$\,dex higher $\xi_\mathrm{ion}(1-f_\mathrm{esc})$, which is explained by their higher H$\alpha$ EW distribution, implying younger ages, higher sSFR and thus more O/B stars. We find significant trends of increasing $\xi_\mathrm{ion}(1-f_\mathrm{esc})$ with increasing H$\alpha$ EW, decreasing UV luminosity, and decreasing UV slope, implying the production of ionizing photons is enhanced in young galaxies with assumed low metallicities. We find no significant evidence for sources with very high ionizing escape fraction ($f_\mathrm{esc}>0.5$) in our sample, based on their photometric properties, even amongst the Lyman-alpha selected galaxies.}
    {This work demonstrates that considering the full distribution of $\xi_\mathrm{ion}$ across galaxy properties is important for assessing the primary drivers of reionization.}


\maketitle

\section{Introduction} \label{sec:intro}

In recent years, we have obtained increasing evidence that the reionization of hydrogen happened fairly late, approximately one billion years after the Big Bang ($z\sim5.5-10$), with a mid-point around $z\sim7-8$ \citep[e.g.,][]{Fan2006,Stark2010,McGreer2015,Mason2018,Davies2018b,Qin2021b,Planck2018,Bolan2022}. However, there is evidence for significant star formation before this time \citep[e.g.,][]{Oesch2018a,Hashimoto2018,McLeod2021} thus it appears that reionization lags behind galaxy formation. The reason for this lag is unknown: we are still lacking a full physical understanding of the reionization process. In particular, we still do not know which types of galaxies drive the process, i.e. which physical mechanisms mediate the production and escape of ionizing photons from galaxies.
In order to produce such a late and fairly rapid reionization, the ionizing population could have been dominated by low mass, UV-faint galaxies with a low ($\sim5\%$) average escape fraction \citep[e.g.,][]{Mason2019c,Qin2021b}. Alternatively, rarer more massive galaxies with higher escape fractions could have been responsible \citep[e.g.,][]{Sharma2017b,Naidu2020}. With only measurements of the timing of reionization, these scenarios are degenerate, thus physical priors on the ionizing properties of galaxies across cosmic time are necessary to pinpoint the sources of reionization.

The total ionizing output of galaxies can be simply parameterized \citep[e.g.,][]{Madau1999,Robertson2010c} as the product of the production rate of ionizing photons relative to non-ionizing UV photons, \xiion \citep[determined by the stellar populations, e.g.,][]{Stanway2016}, and the fraction of ionizing photons which escape the interstellar medium (ISM) into the intergalactic medium (IGM), \fesc \citep[determined by the structure and ionization state of the ISM, likely shaped by star formation and feedback, e.g.,][]{Trebitsch2017,Ma2020}. Both of these quantities are also expected to vary with time in an individual galaxy, for example due to the lifetime and properties of young stellar populations, and the effects of feedback and bursty star formation on the ISM. 

While we can easily observe the non-ionizing UV photons from galaxies, the high optical depth of the IGM to ionizing photons makes direct measurements of the escaping ionizing spectrum statistically unlikely at $z\simgt3$ \citep{Inoue2014, Becker2021,vanzella_2018}. Alternatively, fluxes of non-resonant recombination lines, emitted by gas which was ionized in HII regions around massive stars, can crucially measure the flux of ionizing photons which do not escape galaxies. In particular, H$\alpha$ emission can be used to directly estimate $(1-\fesc)\xiion$ \citep[e.g.,][]{Leitherer1995,Bouwens2016,Shivaei2018,Emami2020}. As \fesc is inferred to be low ($\simlt10\%$) on average for Lyman-break galaxies at $z\sim2-4$ \citep{Steidel2018,Begley2022,Pahl2022} measurements of H$\alpha$ should trace the intrinsic production of ionizing photons reasonably well. \xiion can also be inferred from the strength of \OiiiHb emission, though due to the dependence of [OIII] emission on metallicity and ionization parameter, the correlation is not as tight as with \Ha \citep[e.g.,][]{Chevallard2018}.

Previous work at $z\simlt2.5$ , where direct H$\alpha$ spectroscopy has been possible from the ground, has found a mean $\log_{10}\xiion [\mathrm{erg\,Hz}^{-1}] \approx 25.3$, with a scatter of $\sim0.3$\,dex, likely dominated by variations in stellar populations between galaxies \citep[e.g.,][]{Shivaei2018,Tang2019}. At higher redshifts where H$\alpha$ redshifts into the infra-red, broad-band photometry with Spitzer has been used extensively to estimate H$\alpha$ line fluxes \citep[e.g., ][]{Schaerer2009,Shim2011,Stark2013, Smit2015,Bouwens2016,Lam2019a,Maseda2020,Stefanon2022c}.

However, due to the limited spatial resolution and sensitivity of Spitzer, previous works had been limited to studying \xiion in isolated, bright ($>L^*$) galaxies where deblending IRAC photometry was possible \citep[e.g,.][]{Bouwens2016} and stacks for fainter galaxies \citep[e.g.,][]{Lam2019a,Maseda2020}. With JWST it is finally possible to extend these studies to individual UV-faint galaxies \citep{Endsley2022}, and obtain rest-frame optical spectroscopy at $z>3$ \citep[e.g.,][]{Sun2022,Williams2022}.

Results from previous analyses have been intriguing but require further investigation. Using stacked IRAC photometry \citet{Lam2019a} found no significant evidence for a strong correlation of \xiion with \MUV. However, \citet{Maseda2020} found a  population of extremely UV-faint  ($\MUV > -16$) galaxies selected as \lya emitters in deep MUSE observations, which have very elevated \xiion compared to higher luminosity galaxies and at fixed H$\alpha$ EW, implying these efficient ionizing galaxies are particularly young and low metallicity. It is thus important to examine the distribution of \xiion at low UV luminosities, and to compare galaxies with and without \lya emission to better understand the demographics of the ionizing population.

Furthermore, using early JWST NIRCam data \citet{Endsley2022} discovered a population of UV-faint ($\MUV \sim -19$) galaxies at $z\sim6.5-8$ with high sSFR but low EW [OIII]+H$\beta$ inferred from photometry. The high sSFR would imply high \xiion due to the increased abundance of O and B stars. To explain the low [OIII]+H$\beta$ EW, \citet{Endsley2022} suggest either these galaxies have extremely low metallicities (reducing oxygen abundance), or alternatively, all nebular lines are reduced. A reduction in all nebular lines could be due to either them being produced in density-bounded HII regions with very high ionizing escape fraction \citep[e.g.][]{Zackrisson2013,Marques_Chaves_2022}, or recent cessation of star formation. At $z\sim3-7$ both [OIII]+H$\beta$ and H$\alpha$ are visible in NIRCam photometry, enabling us to test these scenarios.

In this paper we make use of deep multi-band HST/ACS, WFC3 and JWST/NIRCam imaging with overlapping MUSE observations, enabling us to blindly detect a spectroscopic sample with precision rest-frame ultra-violet to optical photometry. We measure the distribution of \xiion over a broader luminosity range ($-23 \simlt \MUV \simlt -15.5$) than previously possible in individual galaxies, due to the excellent resolution and sensitivity of NIRCam at rest-optical wavelength compared to Spitzer/IRAC, and the power of gravitational lensing. We explore correlations of \xiion with empirical galaxy properties. We find significant trends of increasing \xiion with decreasing UV luminosity, UV $\beta$ slope and with increasing \Ha EW, all implying the strongest ionizers are  young sources with expected low metallicities. We also explore whether our sample shows evidence for very low metallicities or extremely high escape fraction.

The paper is structured as follows. In Section~\ref{sec:data} we describe the photometric and spectroscopic data for our study. In Section~\ref{sec:xiion} we describe how we infer the ionizing production rate \xiion and we describe our results of correlations between \xiion and other galaxy properties, and comparison to the literature in Section~\ref{sec:res}. We discuss our results and state our conclusions in Section~\ref{sec:conc}.

We assume a flat $\Lambda\mathrm{CDM}$ cosmology with $\Omega_m=0.3,\,\Omega_\Lambda=0.7,\,h=0.7$, all magnitudes are in the AB system.

\section{Data} \label{sec:data}

For this work we select fields with multi-band HST/ACS and JWST/NIRCam imaging and overlapping MUSE spectroscopy. We select sources detected with \lya emission ($z\sim2.9-6.7$ in MUSE) and sources with high probability of being in the same redshift range based on photometric redshift, and use the HST + JWST photometry to extract optical emission line fluxes. Below we describe the datasets and the selection of our sample.

\subsection{Imaging} \label{sec:data_jwst}

We use JWST NIRCam imaging in parallel to and of the cluster Abell 2744 from the GLASS-JWST program ERS-1324 \citep[PI Treu][]{Treu2022} and UNCOVER\footnote{\url{https://www.stsci.edu/jwst/science-execution/program-information.html?id=2561}} program GO-2561 (co-PIs Labbé, Bezanson).

The GLASS-JWST NIRCam observations discussed in this paper were taken in parallel to NIRISS observations of the cluster Abell 2744 on June 28-29 2022. They are centered at RA$=3.5017025$ deg and Dec$=-30.3375436$ deg, and consist of imaging in seven bands: F090W (total exposure time: 11520 seconds), F115W (11520 s.), F150W (6120 s.), F200W (5400 s.), F277W (5400 s.), F356W (6120 s.), and F444W (23400 s.). The UNCOVER NIRCam observations of the Abell 2744 cluster were taken on November 2-15 2022. They are centered at RA=$3.5760475$\,deg and Dec$=-30.37946$\,deg and consist of imaging in seven bands: F115W (10823 s.), F150W (10823 s.), F200W (6700 s.), F277W (6700 s.), F356W (6700 s.), F410M (6700 s.) and F444W (8246 s.).

In our analysis, we also include new and archival HST imaging, from which ACS imaging is particularly important for constraining photometric redshifts. This includes new HST/ACS data in F606W (59530 s.), F775W (23550 s.), and F814W (123920 s) from HST-GO/DD program 17231\footnote{\url{https://www.stsci.edu/cgi-bin/get-proposal-info?id=17231\&observatory=HST}} (PI Treu), as well as archival data acquired under the Hubble Frontier Fields program \citep[HST-GO/DD-13495, PI Lotz,][]{Lotz2017}, BUFFALO \citep[HST-GO-15117, PI Steinhardt,][]{Steinhardt2020} and programs HST-GO-11689 (PI Dupke), HST-GO-11386 (PI Rodney), HST-GO-13389 (PI Siana), HST-GO-15940 (PI Ribeiro) and HST-SNAP-16729 (PI Kelly). Not all HST bands cover every object in our sample, but we only keep objects in our sample that have a well-constrained photometric redshift, usually meaning that there is ACS coverage (see Section \ref{sec:data_sample}). We also include HST/WFC3 imaging for completeness, but these are generally not as constraining as the NIRCam fluxes. 

The image reduction and calibration, and the methods used to detect sources and measure multi-band photometry in both fields closely follow that of Brammer et al., (in prep). Briefly, we pull calibrated images from the MAST archive\footnote{\url{https://archive.stsci.edu}} and process them with the \texttt{grizli} pipeline \citep{Brammer2022}. The pipeline first aligns the exposures to external catalogs and to each other and corrects for any distortion within the image. Following this, we subtract a sky-level background, divide out flat-field structure using custom flat-field images, and correct for $1/f$ noise. We also correct for NIRCam image anomalies, which include persistence, any remaining cosmic rays, and `snowballs' (see \citealp{Rigby2022}). Finally, we apply zeropoint corrections calculated by G. Brammer\footnote{\url{https://github.com/gbrammer/grizli/pull/107}} and drizzle all exposures to a common pixel grid. 

For source detection, we use SEP \citep{Barbary2018} to perform aperture photometry on the F444W detection image in each field. 

\subsection{VLT/MUSE spectroscopy} \label{sec:data_muse}

MUSE spectroscopy in the Abell 2744 cluster were obtained through ESO program 094.A-0115 (PI Richard) and are described by \citet{Mahler2018} and \citet{Richard2021}. We use their publicly available catalog to select \lya emitting galaxies. The data comprise a 4 sq. arcmin mosaic centered on the cluster core. Four 1 sq. arcmin quadrants were observed for a total of 3.5, 4, 4 and 5 hours respectively, and the center of the cluster was observed for an additional 2 hours. The median line flux 1$\sigma$ uncertainty in the MUSE data is $3.6\times10^{-19}$\,erg s$^{-1}$ cm$^{-2}$. This corresponds to an 5$\sigma$ EW limit of $\sim4-30$\,\AA\ over $z\sim3-7$ for a galaxy with $\MUV = -19$ (the median for our sample, before accounting for magnification, as EW is invariant under magnification).

VLT/MUSE spectroscopy in the GLASS-JWST NIRCam fields were obtained through a new ESO DDT program 109.24EZ.001 (co-PIs Mason, Vanzella) on the nights of July 28 and August 20 2022. The data comprise 5 pointings (4 pointings -- 4 sq. arcmin -- overlapping with NIRCam imaging) each with 1 hour exposure time. The raw data are publicly available on the ESO archive\footnote{\url{http://archive.eso.org/wdb/wdb/eso/sched_rep_arc/query?progid=109.24EZ.001}}. The reduction, calibration and source detection methods used for this work are identical to techniques described in previous works \citep{Caminha2017,Caminha2019}. A full assessment of the depth is ongoing but based on the $\sim4$\,hour depth of the \citet{Mahler2018} observations described above, we estimate a 5$\sigma$ EW limit of $\sim8-60$\,\AA\ in these shallower data.

In this work we use 102 spectroscopic confirmations at $z\sim2.9-6.7$: 42 from the GLASS-JWST NIRCam fields and 60 from the Abell 2744 cluster field.

\subsection{Gravitational lensing magnification} \label{sec:data_lensing}

For the galaxies detected in the core of the Abell 2744 cluster we correct for gravitational lensing magnification using the model by \citet{Bergamini2022}. The median magnification of the sample is $\mu=3.54$ with 90\% of the galaxies having $\mu = 2 - 20$. We remove sources with a magnification with $\mu>50$ (12 sources) due to high uncertainties in the model near the critical curves. The galaxies in the parallel fields are $\sim3-10^\prime$ away from the cluster core where the magnification is expected to be modest ($\mu\approx1$), we do not account for magnification of those sources.

\subsection{Sample selection} \label{sec:data_sample}

For this work, we focus on selecting a sample of galaxies at $z\sim3-7$ with high purity. We select 102 MUSE \lya-detected galaxies with overlapping HST/ACS and JWST/NIRCam data as described above. We also select a comparison sample of galaxies based on peak photometric redshift, within the same footprint as the MUSE observations, which we expect to have slightly lower \Ha EW than the \lya selected sample.

We find the photometric redshift distribution of all sources detected as described in Section~\ref{sec:data_jwst} using EAZY \citep{Brammer2008}, given all available photometric bands. To build a photometric sample with high purity, following \citet{Bouwens2016}, we select sources with the peak of their photometric redshift between $2.9 < z < 6.7$, and keep only sources which have 90\% of the redshift probability density between $\Delta z \sim 1$ of the peak of their distribution. The resulting high purity photometric sample consists of 268 galaxies.

\begin{figure}
\includegraphics[width=\columnwidth]{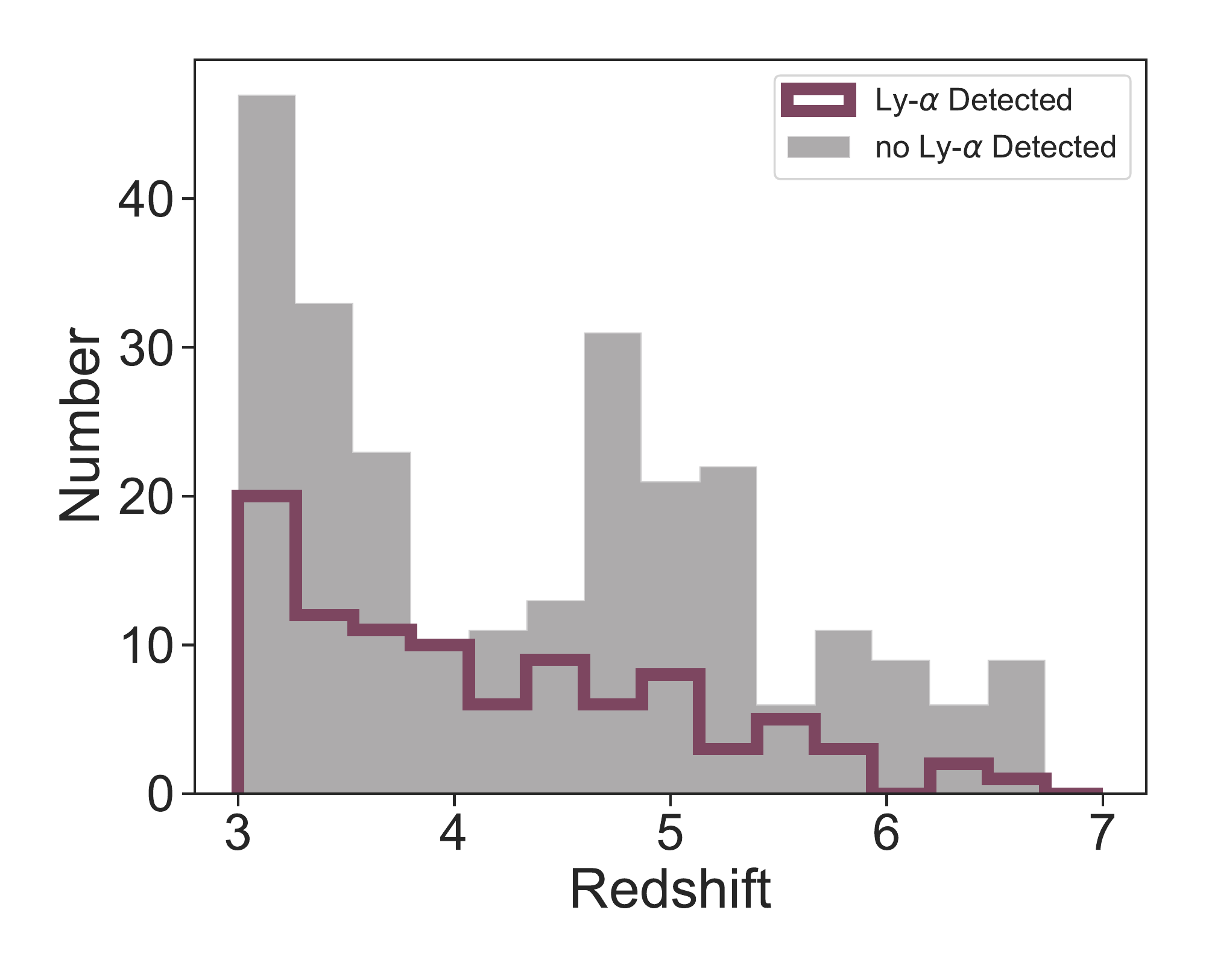}
\includegraphics[width=\columnwidth]{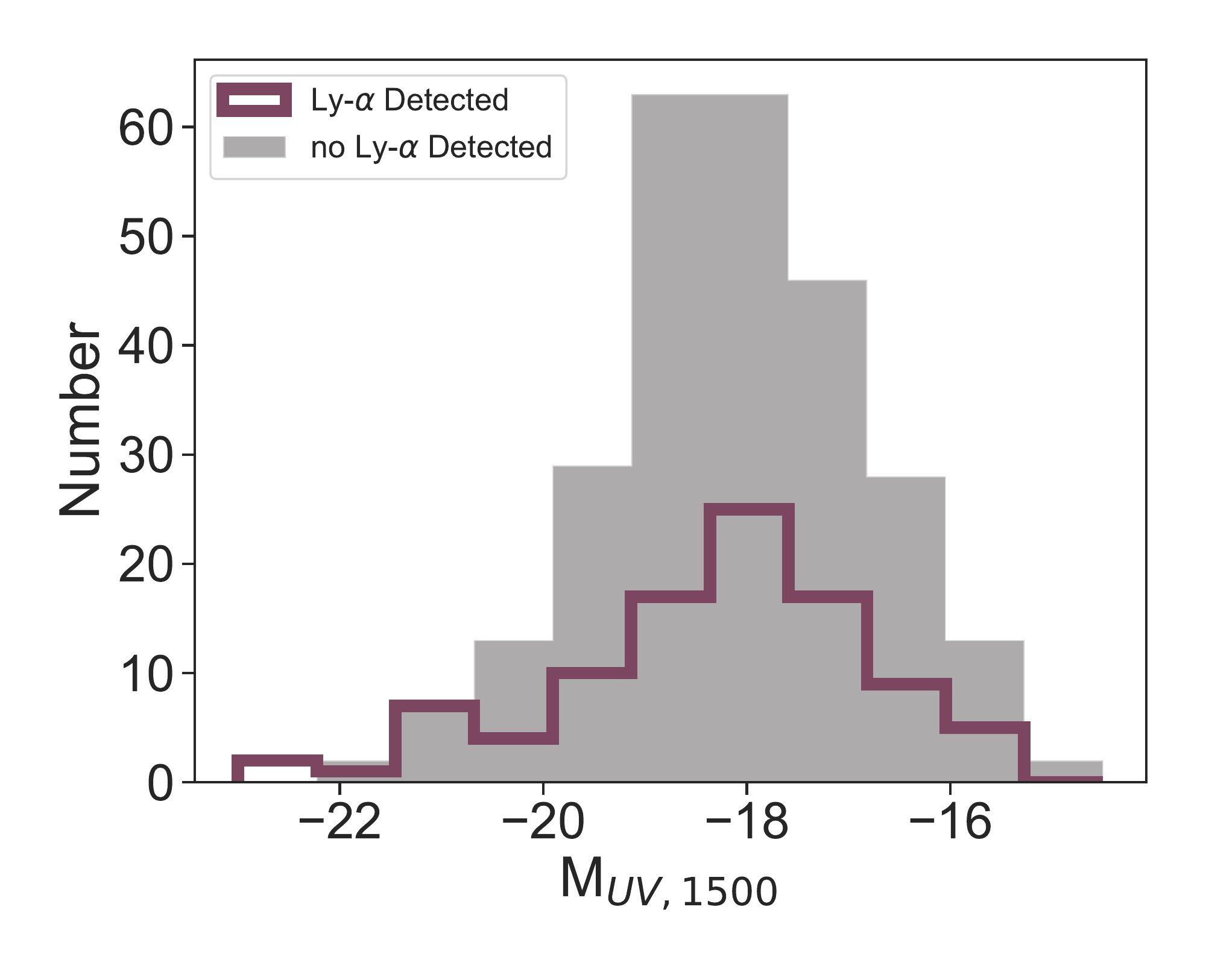}
\caption{Galaxies studied in this work; in purple \lya detected galaxies, and in gray galaxies photometrically selected (with no \lya detected). \textit{Top:} Distribution of redshifts for the spectroscopic and photometric samples. We show the spectroscopic redshift where available, or the peak photometric redshift. \textit{Bottom:} UV magnitude distribution for our sample, we find a median value of $-18.14\pm1.58$, with no statistically significant difference between both samples.
\label{fig:Muv_dist}
}
\end{figure}

The redshift and UV magnitude distribution of our sample is shown in Figure \ref{fig:Muv_dist}. The median redshift of the full sample is 4.02, with the \lya-selected sample having median redshift 3.95. The median \MUV is -18.1, with a Kolmogorov-Smirnov (KS) test showing no significant difference between the \lya and photometrically selected samples.

\section{Inferring the ionizing photon production rate} \label{sec:xiion}

\subsection{Inferring nebular emission line strengths from photometry} \label{sec:xiion_SED}

To estimate nebular emission line fluxes from broad-band photometry, we follow approaches in the literature and fit the spectral energy distribution (SED) to the full photometry, excluding bands we expect to contain strong nebular emission lines \citep[e.g.,][]{Shim2011,Stark2013,Marmol-Queralto2015,Bouwens2016}. This provides us with a model for the continuum flux in those bands which we can subtract from the observed photometry to infer the line flux.

We use BAGPIPES to fit SEDs \citep{Carnall2018}. We adopt BC03 \citep{Bruzual2003} templates, and exclude any nebular emission contribution. We do not consider any broadbands where \Ha or \OiiiHb are observed according to each galaxy's redshift. For ease of comparison to the literature (e.g., \citealp{Maseda2020,Lam2019a}), we assume a \citet{Chabrier2003} Initial Mass Function and an Small Magellenic Cloud (SMC, \citealp{Prevot1984}) dust attenuation law, allowing $A_V$ to vary from $0-3$ mag. Because metallicity is not well known at the range of redshifts we explore, we allow metallicity to vary from $0-2\,Z_{\odot}$. Because star formation histories are notoriously difficult to constrain at high-redshift \citep{Strait2021}, we assume an exponentially rising delayed $\tau$ star formation history, allowing $\tau$ to vary freely. For the spectroscopically confirmed Lyman $\alpha$ emitters, we fix the redshift at the \lya redshift. For our photometric sample, we use the photometric redshift obtained from EAZY with uniform prior with $\Delta z = 1$. (see Section~\ref{sec:data_sample}).

We then compare  the SED model of the galaxy's continuum to the broadbands where \Ha or \OiiiHb fall. We multiply the non-nebular SED posteriors with the transmission of the aforementioned broadbands to obtain the contribution of the galaxy's continuum to the observed flux. By subtracting this continuum flux contribution to the observed photometry, we then are able to recover the flux distributions of \Ha and \OiiiHb emission lines for each galaxy. We compared our measurements with a sample of 6 galaxies with \OiiiHb EW measurements from the GLASS-ERS program using JWST/NIRISS \citep{Boyett2022b}, finding that our method recovers the EW of these sources within $\sim20-40$\%. A full comparison of these photometric inference methods is left to future work.

There are some limitations to our method for obtaining line fluxes, such as contamination from the 4000-\AA\ break in the broadband containing \OiiiHb, the chance that the line falls outside the effective width of any of our broadband filters, or \Ha and \OiiiHb falling on the same band. We consider a contribution of 6.8\% from [NII] to the calculated \Ha flux, and 9.5\% from [SII] according to \citet{Anders_2003}. We remove galaxies with a poor $\chi ^2$ ($>$50) score  on their SED fit, we choose these value by ignoring all galaxies on the high-end of the $\chi ^2$ distribution. 

The advantage of this approach, unlike estimating line fluxes directly from the SED fitting, is that it does not depend strongly on star formation history assumptions and allows us to make a mostly empirical measurement of the line fluxes. We obtain comparable results using the flux in the band red-ward of \Ha as the continuum flux, assuming a flat optical continuum \citep[see also e.g.,][]{Maseda2020}. Estimating other physical parameters such as star-formation rate and stellar mass from the SED fitting did not give reliable results, since it was too dependent on the initial assumptions, and needed extremely young ages (<10Myrs) and an instantaneous burst of star-formation to recreate the observed nebular emissions..

The following results will consist of: 83 and 64 \lya-emitting galaxies with \Ha and \OiiiHb emission line measurements respectively, a photometric sample of 220 and 203 galaxies with \Ha, and \OiiiHb emission line measurements respectively. We see both lines in 62 \lya galaxies and 177 photometrically selected galaxies.  We see no apparent biases in our \MUV distribution after narrowing down the sample. Nebular emission flux errors are derived from the 68\% confidence interval of the resulting distributions.

\subsection{Measuring UV absolute magnitude and slope} \label{sec:Muv_Slope}

To infer the UV absolute magnitude, \MUV (magnitude at 1500\AA) and $\beta$ slope we fit a power law \citep[e.g.,][]{Rogers2013} $f_\lambda\propto\lambda^\beta$ to the fluxes from the HST and JWST bands.
We perform the fit using a Markov Chain Monte Carlo sampling using the python module \texttt{emcee} \citep{Foreman-Mackey2013}. We assume flat priors for $\beta$ and \MUV, with bounds $-4< \beta <1$ and $-25< M_{UV} < -12$, sufficient to explore the common value ranges for galaxies \citep[e.g.,][]{Bouwens2014a}.

To obtain the photometric bands that are observing the UV-restframe of our galaxies, we exclude any bands that fall bluewards from Lyman-$\alpha$ that might be affected by the Lyman-Break. For the same reason, we exclude bands redwards from the 4000\AA\ break in the restframe. After these requirements, we are left with between 3 and 4 bands for each source. In the case of galaxies with Lyman-$\alpha$ detected in MUSE, we use the line's redshift. For photometrically-selected galaxies, in each call of the likelihood, we randomly draw a redshift from a Normal distribution $N(\mu = z_{phot},\sigma=0.5)$, and select the appropriate photometric bands.  For lensed sources, we consider magnifications and apply them following the same random draw method as redshift. We use the corresponding magnification and error obtained from the \citet{Bergamini2022} lensing model.

\subsection{Determination of $\xi_\mathrm{ion}$} \label{sec:xiion_calc}

We define the production rate of ionizing photons, $\xiion$. This is given by the ratio between the luminosity of observed ionizing photons and the intrinsic luminosity of the ionizing UV photons \citep[e.g.,][]{Leitherer1995}:
\BE \label{eqn:xi_ion}
\xiion = \frac{L_{\mathrm{H}\alpha}}{(1-\fesc) L_{\mathrm{UV},\nu}^\mathrm{intr}} \times 7.37\times10^{11} \; \mathrm{Hz\,erg}^{-1}
\EE
where $L_{\mathrm{H}\alpha}$ is the unattenuated H$\alpha$ luminosity in erg s$^{-1}$ and $L_{\nu,\mathrm{UV,intr}}$ is the intrinsic UV luminosity density at 1500\AA. The models from where the conversion factor is derived, assume a young population of massive stars equivalent to a massive HII region. We assume this type of environment to be similar to what we would find in young galaxies.

Because H$\alpha$ is produced by the excitation of hydrogen gas from ionizing radiation that \textit{does not escape} the galaxy, and we cannot measure \fesc directly in our sample, we note that we are obtaining the production rate of ionizing photons which did not escape the galaxy, $\xiion(1 - \fesc)$.

We first calculate $L_\mathrm{_{H\alpha}}$ directly from the SED obtained in Section \ref{sec:xiion_SED}, after accounting for dust attenuation \citep{Prevot1984}.
To obtain the intrinsic value of the UV luminosity, we take into account the dust attenuation following \citet{Lam2019a}, where the intrinsic UV luminosity is defined as $
L_{\mathrm{UV},\nu}^\mathrm{intr} =L_{\mathrm{UV},\nu} / f_\mathrm{esc,UV}$. $f_\mathrm{esc,UV}$ is the fraction of escaping UV photons not absorbed by the dust. For this, we use the Small Magellanic Cloud dust law defined by \citet{Prevot1984}:
\BE \label{eqn:dust_law}
f_\mathrm{esc,UV} = 10^{-1.1(\beta+2.23)/2.5} , \quad \beta > -2.23
\EE
Where $\beta$ is the UV slope obtained in Section \ref{sec:Muv_Slope}. Galaxies with slopes bluer than $\beta <$ -2.23 are assumed dust-free and we do not correct for dust.

In the following, uncertainties on $\xiion$ are the 68\% confidence intervals, obtained from propagating the uncertainty in the H$\alpha$ flux from its resulting distribution as described in Section~\ref{sec:xiion_SED}, and the posterior distributions for $\beta$ and \MUV as described in Section~\ref{sec:Muv_Slope}.

\subsection{Correlation Analysis}
\label{sec:xiion_corr}

For the purpose of studying the correlations between galaxy properties we use the python package \texttt{linmix}\footnote{\url{https://github.com/jmeyers314/linmix}} to do Bayesian linear regression including intrinsic scatter and accounting for two-dimensional errors \citep{Kelly2007}. We fit for $\log_{10}[(1-\fesc)\xiion]= \alpha + \beta X + \epsilon$, where $\epsilon$ is the intrinsic scatter which is assumed to be normally distributed with variance $\sigma_{\epsilon}^2$. We recover the best fit trend line from the posteriors, as well the 68\% confidence interval on the parameters. We report the results in Table \ref{tab:fit} and show the best-fit line on plots.

\section{Results} \label{sec:res}
In the following section we present our results. In Section~\ref{sec:res_xiion} we study the trends between \xiion, \Ha EW, \MUV, and $\beta$ slope, and in Section~\ref{sec:res_fesc_Z} we investigate whether our sample shows evidence for very high ionizing photon escape fraction and/or very low metallicity galaxies. 

\begin{figure}
\includegraphics[width=\columnwidth]{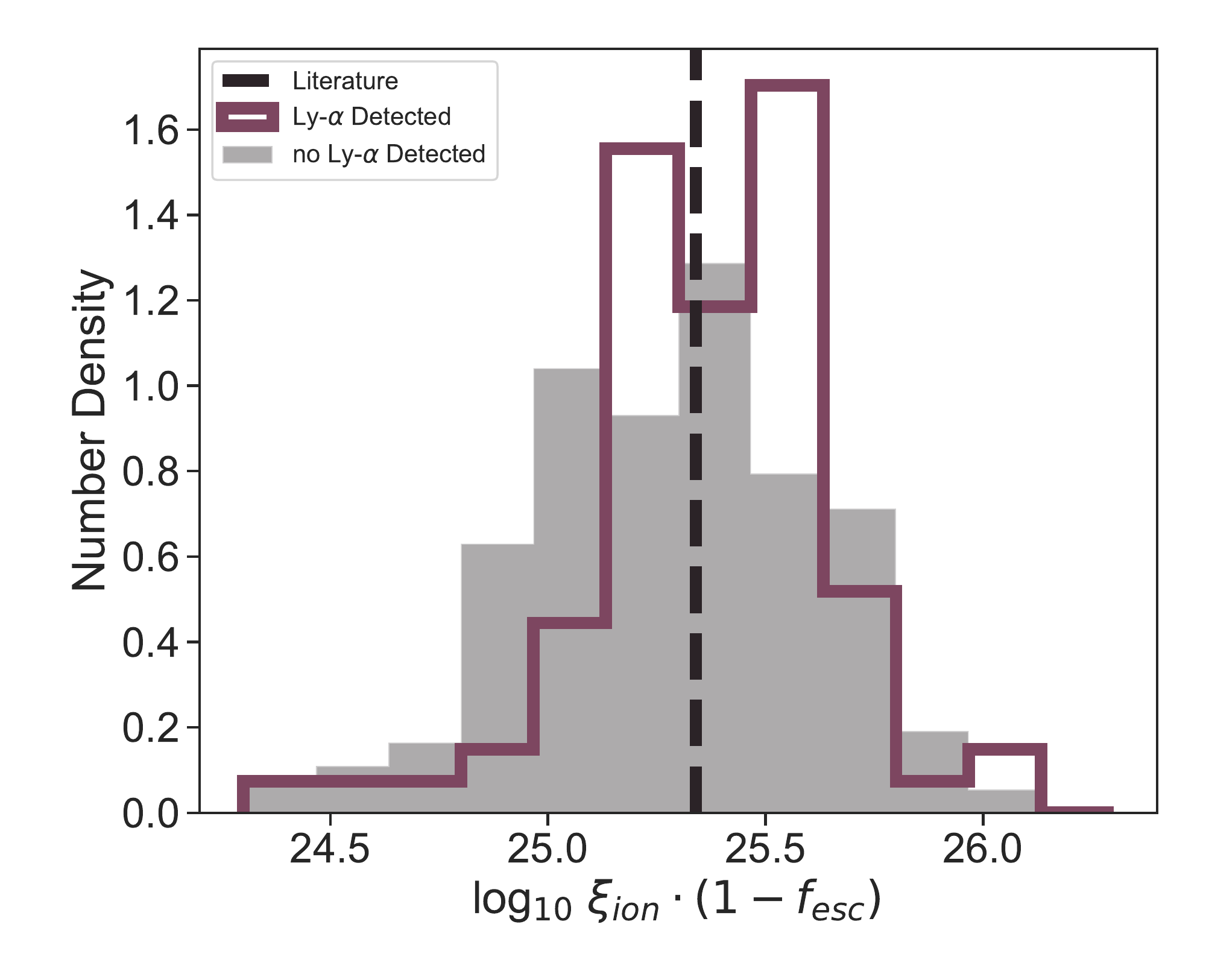}
\caption{Distribution of (1-\fesc)\xiion. The purple histogram includes galaxies with; \lya emission detection, and in grey galaxies without \lya detected. Overall, \lya emitting galaxies show stronger ionizing photon production than galaxies with no \lya emission selected galaxies, with median values 25.39$\pm$0.64 and 25.31$\pm$0.43 respectively. We show the median relation from the literature at $z\sim2-5$ as a dashed black line \citep[e.g.,][]{Shivaei2018,Bouwens2016,Lam2019a}
\label{fig:xi_ion_pdf}
}
\end{figure}

\begin{figure}
\includegraphics[width=\columnwidth]{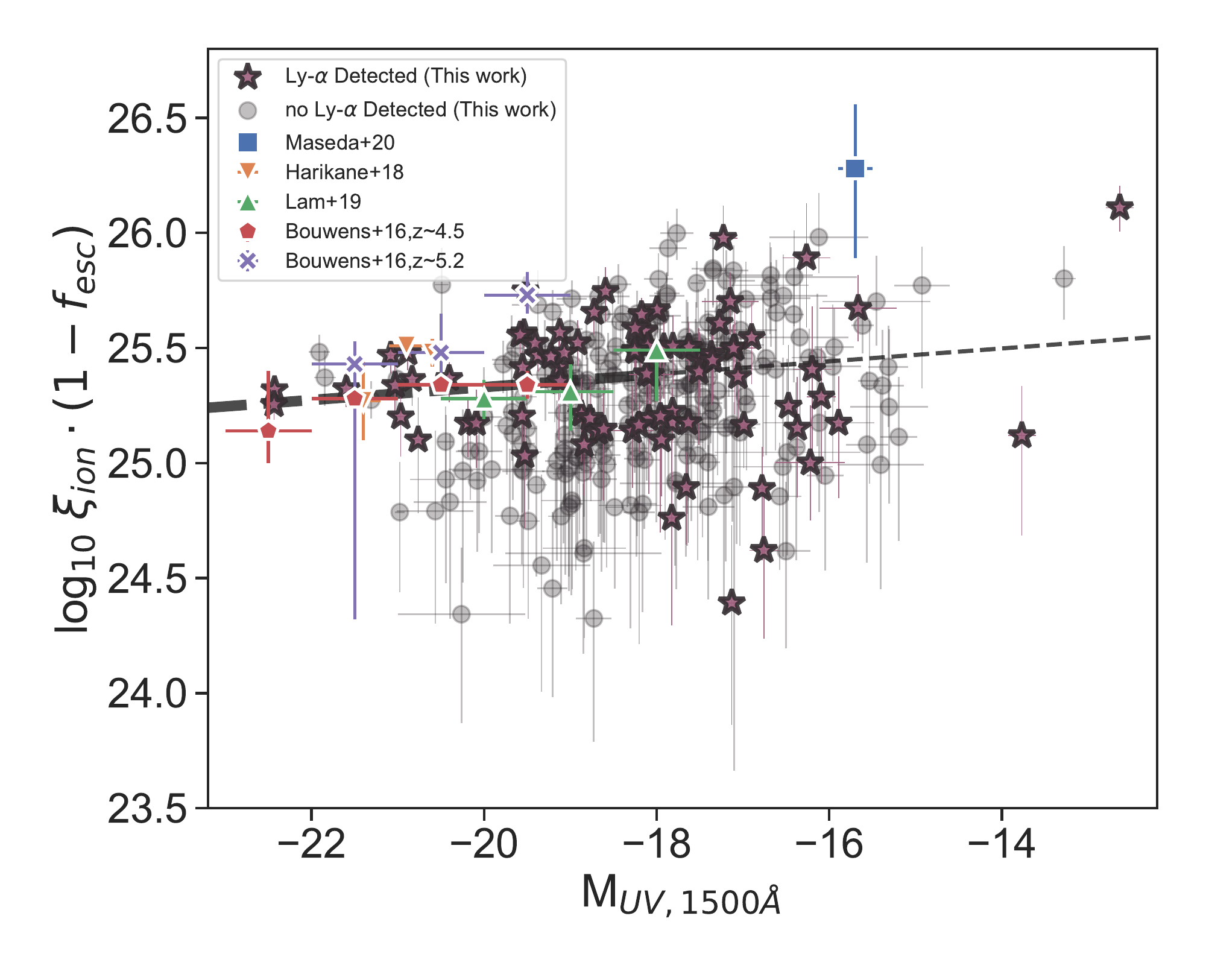}
\caption{\MUV vs (1-\fesc)\xiion. In purple circles we show \lya detected galaxies. In gray circles are the photometrically selected sample with no \lya detected. We show data from \citet{Maseda2020,Harikane2018,Lam2019a} and \citet{Bouwens2016} as colored boxes for comparison. We find evidence for an increase in $\log_{10}(1-\fesc)\xiion$ towards fainter UV magnitude, with a slope of $0.03\pm0.02$, only considering the range where our sample is \MUV complete ($\MUV<-18.1$). We show literature constraints at similar redshifts as colored shapes \citep{Bouwens2016,Harikane2018,Lam2019a,Maseda2020}, noting that all constraints fainter than $\MUV \simgt -20$ were obtained by stacking IRAC photometry.
\label{fig:xi_ion_Muv}
}
\end{figure}

\begin{figure*}[htb]
\center
\includegraphics[width=\textwidth]{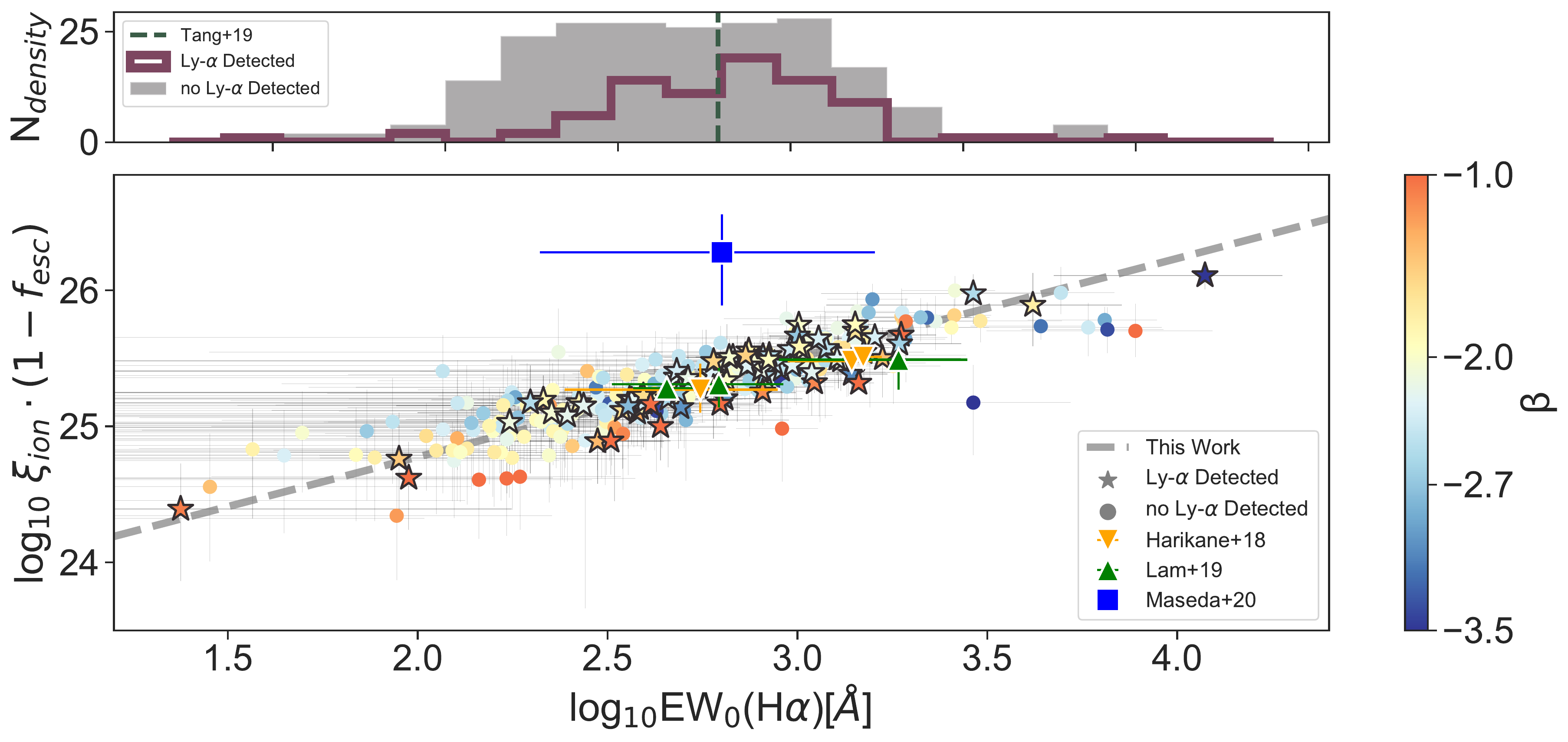}
\caption{We compare \Ha equivalent width with the ionizing photon production that does not escape the galaxy. As stars we show \lya detected galaxies, as circles are photometrically selected galaxies with no \lya. As above, error bars are only shown for 30\% of the sources for clarity. We color code these two samples by UV $\beta$ slope. On top we show the distribution of \Ha EW for the same two samples, compared to the values found by \citet{Tang2019}. We add data from \citet{Harikane2018} and \citet{Lam2019a} for comparison, which is at the high end of our observed \Ha EW distribution. We see that stronger \xiion very strongly correlates with \Ha EW. Galaxies with detected \lya emission have an \Ha EW distribution with higher values, median $732\pm187$\,\AA\ compared to $457\pm161$\, with a Kolmogorov–Smirnov test p-value $\ll0.01$. The sources with the reddest UV slopes lie systematically below the best-fit relation at fixed \Ha EW.
\label{fig:xi_ion_Halpha}
}
\end{figure*}

\begin{figure}
\includegraphics[width=\columnwidth]{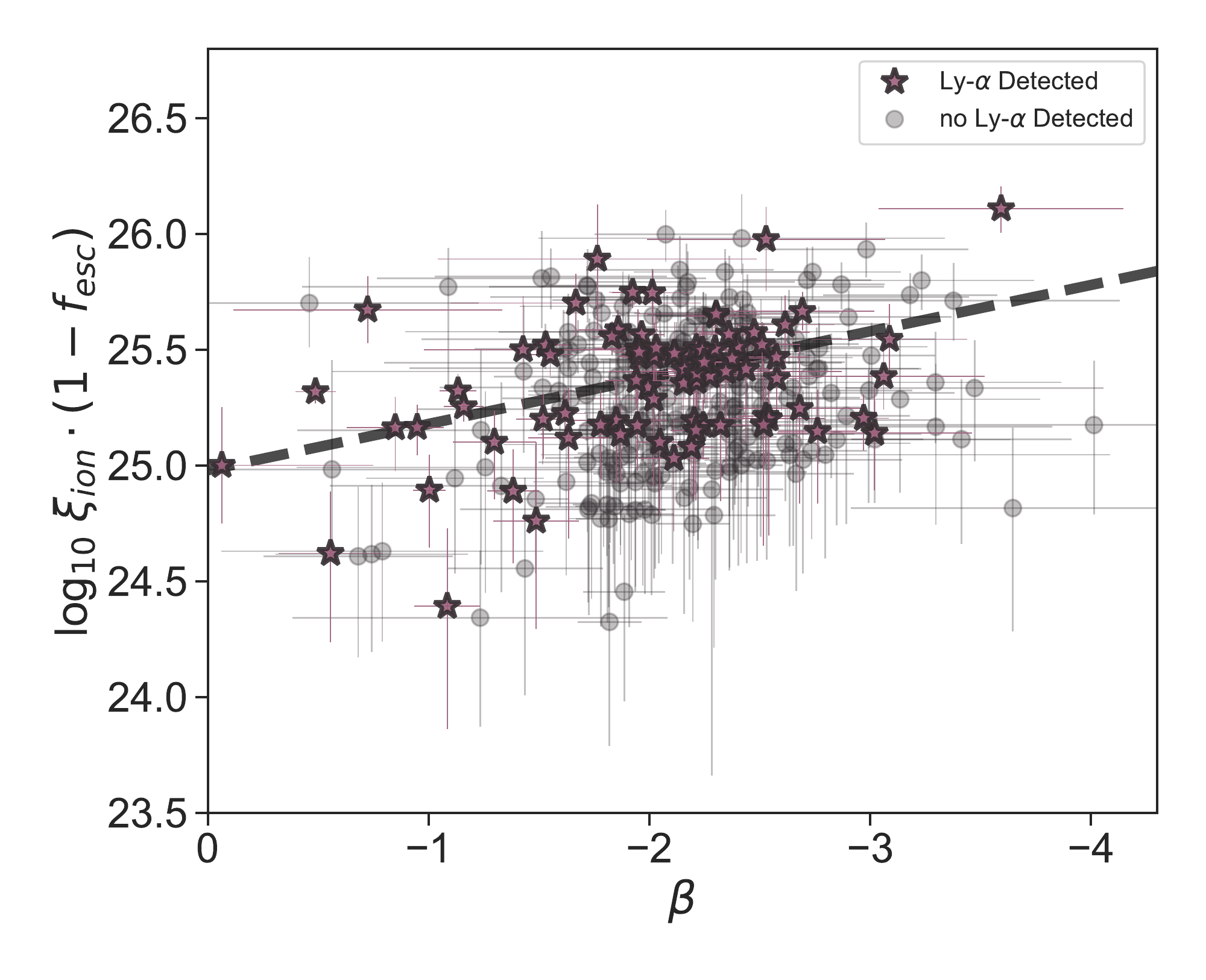}
\caption{UV $\beta$ slope vs \xiion(1-f$_{esc}$). In purple we show \lya detected galaxies. In gray is the photometrically selected sample with no \lya detected. As above, error bars are only shown for 30\% of the sources for clarity. We add the stacked measurements from \citet{Lam2019a} for comparison. We find a very weak trend of increasing \xiion with decreasing $\beta$, with a linear slope of $-0.10\pm0.06$.
\label{fig:xi_ion_beta}
}
\end{figure}

\subsection{Behaviour of \xiion}
\label{sec:res_xiion}

Figure~\ref{fig:xi_ion_pdf} shows the distribution of $(1-\fesc)\xiion$ for our \lya-selected and photometric samples. 
We find median values of $\log_{10} \xiion$ 25.39$\pm$0.64 and 25.31$\pm$0.43 respectively; with 25.33$\pm$0.47 [Hz erg$^{-1}$] for the complete data set. We find an intrinsic scatter of 0.42\,dex, obtained by subtracting in quadrature the average uncertainty in $\log_{10} \xiion$ ($=0.21$\,dex) from the standard deviation of the observed distribution. The recovered intrinsic scatter is broader by $\sim0.1$\,dex than that found by \citet{Bouwens2016} and \citet{Shivaei2018} in $\MUV \simlt -20$ galaxies. The broad distribution of \xiion is likely an outcome of the broad range of stellar populations in these galaxies, i.e. due to a range of star formation histories and thus ages, and stellar metallicity \citep[see e.g.,][]{Shivaei2018}.

We perform a two sample Kolmogorov–Smirnov (KS) test to test whether the \lya-selected and photometric samples are drawn from the same distribution. We recover p-value of 0.03, meaning it is likely that the underlying distributions are different, consistent with the results from \citep{Saldana-Lopez2023}, where a statistically significant difference is found between the \xiion distributions of LAEs and non-LAEs at $z\sim3-5$. Given that galaxies with strong \lya emission also likely have high ionizing photon escape fractions \citep[e.g.,][]{Verhamme2015,Dijkstra2016a} it is likely that the intrinsic ionizing photon production efficiency of these galaxies is even higher than what we can infer based on H$\alpha$ emission.

Figure~\ref{fig:xi_ion_Muv} shows $(1-\fesc)\xiion$ versus UV magnitude and demonstrates the revolutionary capabilities of MUSE and JWST/NIRCam: we are able to spectroscopically confirm extremely UV-faint galaxies via their high \lya EW and we are able to infer \Ha, and therefore \xiion, from much fainter individual galaxies than was previously possible with Spitzer, where stacking was necessary at $\MUV \simgt -20$ \citep[e.g.,][]{Lam2019a,Maseda2020}. We reach $\sim1$\,dex lower than any previous studies at similar redshifts without the need of stacking methods. We can reach individual detections of very faint galaxies, \MUV $<-17$. We also find results consistent with those at $z\sim2$ \citep{Shivaei2018} and at $z\sim4-5$ for $>L^*$ galaxies \citep{Bouwens2016} and $<L^*$ galaxies \citep[][where a stacking analysis was used]{Lam2019a}, as shown in Figure \ref{fig:xi_ion_pdf}. We note our observations demonstrate the large scatter in $(1-\fesc)\xiion$ at fixed \MUV which was not possible to observe in previous analyses which used stacking of Spitzer photometry for UV-faint galaxies. 

As described in Section~\ref{sec:xiion_corr} we perform a linear regression to assess correlations in our data. In contrast to \citet{Lam2019a} we find significant evidence for a weak trend between \xiion and \MUV, where the highest \xiion tends to come from the faintest galaxies. Since our sample is not \MUV complete, we only study the correlation up to the peak of our \MUV distribution (=-18.14) in Figure \ref{fig:Muv_dist}. We find $\log_{10} [(1-\fesc)\xiion] =(0.03\pm0.02)(\MUV+20) + 25.36\pm0.03$, but with large scatter (see Table~\ref{tab:fit}).

Figure \ref{fig:xi_ion_Halpha} shows that \xiion follows a strong trend with \Ha EW, as found in previous work \citep{Harikane2018,Lam2019a,Tang2019}. Previous works were limited to only the highest \Ha EW values, while we reach $\sim 0.75$\,dex lower due to the sensitivity of NIRCam. This trend is consistent with a picture where \xiion is elevated in the youngest, most highly star-forming galaxies \citep[e.g.,][]{Tang2019}. We find $\log_{10} [(1-\fesc)\xiion]= (0.73\pm0.04)(\log_{10} EW_{\mathrm{H}\alpha}-2.5) + 25.15 \pm 0.02$. The measurement by \citet{Maseda2020}, obtained from a stack of extremely UV-faint galaxies with high \lya EW, lies significantly above our sample and the rest of the literature, with higher $(1-\fesc)\xiion$ at fixed H$\alpha$ EW. As discussed by \citet{Maseda2020} this likely implies their sources have a much lower gas-phase metallicity than other samples.

We also find that \lya-selected galaxies have a higher \Ha EW than the photometrically-selected sample, (median EW=$732\pm187$\,\AA\ compared to $457\pm161$\,\AA\ for the photometric sample). A two-sample Kolmogorov-Smirnov test establishes that the EW distributions of the two samples are different (p-value $\ll0.01$). This is likely to be the primary driver of the increased \xiion distribution for the \lya-selected sample (Figure~\ref{fig:xi_ion_pdf}).

At fixed \Ha EW we see a clear tendency for galaxies having very blue $\beta$ UV slopes to have elevated \xiion (Figure \ref{fig:xi_ion_Halpha}). This trend is also seen in the full sample - Figure~\ref{fig:xi_ion_beta}, where we find high \xiion is weakly correlated to blue $\beta$ slope, but with large scatter. We find $\log_{10} (1-\fesc)\xiion = (-0.20\pm0.04)(\beta + 2) + 25.41\pm0.01$ (see Table~\ref{tab:fit}). Similar correlations have been seen at z$\sim$6 \citep[i.e.][]{Ning2023}. Using KS test we find no significant difference in the $\beta$ distributions for the \lya and photometric samples. Our sample has a median $\beta=-2.1$.

\begin{table*}
\caption{\label{tab:fit}Linear fitting parameters for trends with $\log_{10} (1-\fesc)\xiion$}
\centering
\begin{tabular}{lrrr}
\hline\hline
Parameter & Slope, $\alpha$ & Intercept, $\beta$ & Scatter variance, $\sigma_\epsilon^2$\\
\hline
$\MUV+20$ & $0.03\pm0.02$ & $25.33\pm 0.03$ &$0.027\pm 0.006$ \\
$\log_{10}\mathrm{EW}_{\mathrm{H}\alpha} - 2.5$ & $0.73\pm 0.04$ &$25.14\pm 0.02$ & $0.003\pm 0.001$\\
$\beta+2$ & $-0.20\pm0.04$ & $25.38\pm 0.01$ & $0.032\pm 0.005$ \\
\hline
\end{tabular}
\tablefoot{We fit for $\log_{10}[(1-\fesc)\xiion]= \alpha + \beta X + \epsilon$, where $\epsilon$ is the intrinsic scatter which is assumed to be normally distributed with variance $\sigma_{\epsilon}^2$.}
\end{table*}
\subsection{A search for high escape fraction and extremely low metallicity galaxies}
\label{sec:res_fesc_Z}

As well as being a tracer for the ionizing photon production of galaxies, nebular emission lines are also sensitive to the escape fraction. \citet{Zackrisson2013} proposed that in galaxies with very high ionizing escape fraction, one would expect a reduction in nebular emission line strength (H$\beta$ EW $\simlt 30$\,\AA) and extremely blue UV slopes ($\beta < -2.5$) due to the lack of nebular continuum. Early JWST observations have discovered potentially very blue galaxies \citep[][though c.f. \citealt{Cullen2022}]{Topping2022} and galaxies with weak nebular line emission yet high sSFR \citep[via \OiiiHb,][]{Endsley2022}, potentially indicating a population with high ionizing escape fraction. However, the observation of low \OiiiHb line strengths could also be caused by very low gas-phase metallicity (decreasing the strength of [OIII] emission) or a recent turnoff in star formation (which would also decrease all nebular emission lines). Given the redshift range of our sample we can infer both \Ha and \OiiiHb line strengths for 241 galaxies, allowing us to test these scenarios and to search for galaxies with high escape fraction. We obtained the \OiiiHb nebular line fluxes as described in Section \ref{sec:xiion_SED}).

\begin{figure}[htb]
\center
\includegraphics[width=0.5\textwidth]{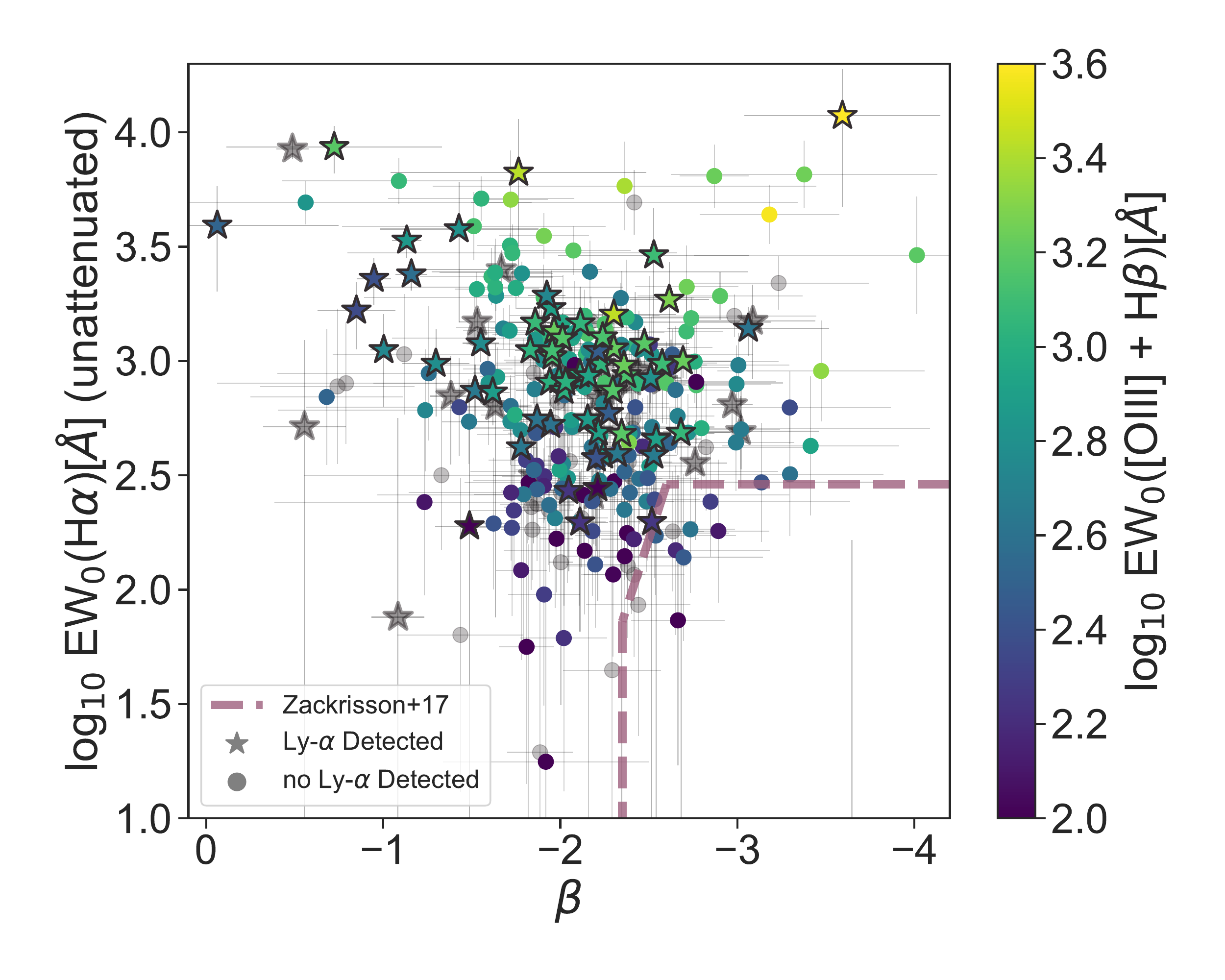}
\caption{Comparison between intrinsic (unattenuated) EW of \Ha and UV $\beta$ slope, color-coded by \OiiiHb EW. \lya galaxies are shown with star shaped markers, and the photometric sample as circles. Galaxies shown in gray, do not have \OiiiHb EW measurements.  We show the region predicted by \citet{Zackrisson2017} to show $\fesc > 0.5$. We rescale from H$\beta$ EW to intrinsic \Ha with a case B recombination scenario of factor 2.89, assuming a flat optical continuum in f$_{\lambda}$, which we confirm from our SED fitting done in \ref{sec:xiion_SED}.
\label{fig:ewha_beta}
}
\end{figure}

In Figure~\ref{fig:ewha_beta} we show UV $\beta$ slopes as a function of intrinsic \Ha EW for our sample (where we correct for dust attenuation as described in Section~\ref{sec:xiion_SED}. We compare our sample to the region proposed by \citet{Zackrisson2017} to have $\fesc > 0.5$. While several sources fall into this region, and also have low \OiiiHb EW ($\simlt 100$\,\AA) the uncertainties are too large to make these robust candidates. We discuss this further in Section~\ref{sec:disc_fesc}.

\begin{figure*}[htb]
\center
\includegraphics[width=\textwidth]{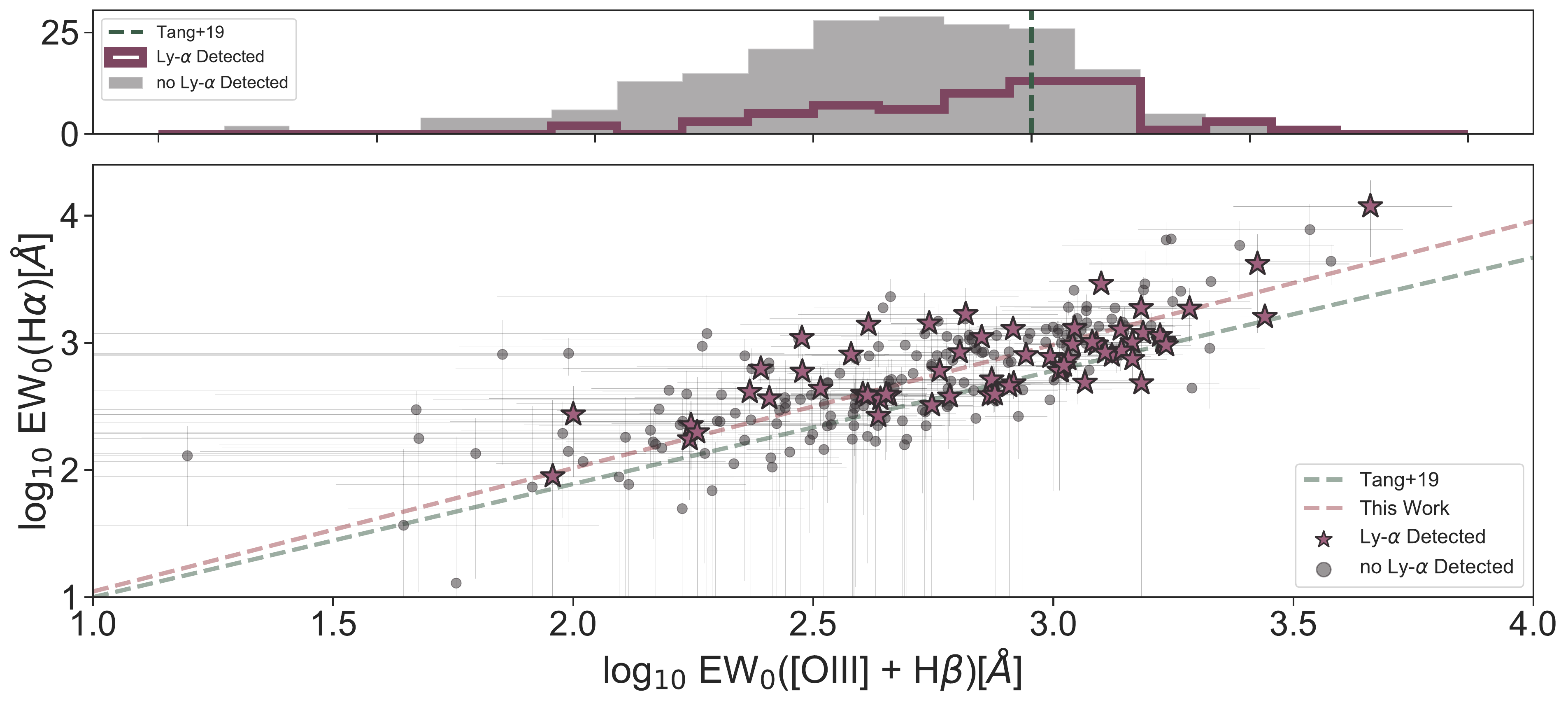}
\caption{Comparison between EW of \Ha and \OiiiHb. We show \lya detected galaxies as stars and the photometrically selected sample with no \lya detected as circles. On top we show the distribution of \OiiiHb EW for both of our samples. We find a very strong correlation between \Ha EW and \OiiiHb EW, though with large scatter. The dashed lines are the correlation trends found for this work (red) and \citet{Tang2019} (green). We find higher \Ha EW/\OiiiHb EW than the $z\sim2$ sample from \citet{Tang2019}, which was selected to have strong [OIII] EW, implying that we might be observing lower metallicity galaxies. 
\label{fig:ewha_ewhb}
}
\end{figure*}

Figure \ref{fig:ewha_ewhb} shows \Ha EW as a function of \OiiiHb for our sample. We see the expected positive correlation between both nebular emission lines, as these lines are all generated by the effects of stellar ionizing radiation. We see a very large scatter (with a range $\sim1.5$\,dex) as expected due to variations in metallicity, temperature, and ionization parameter which will affect the strength of [OIII] individual galaxies \citep[e.g.,][]{Maiolino2008,Steidel2014,Sanders2021}. We find $\log_{10} EW(\Ha) = 0.97\pm0.06(\log_{10} EW(\OiiiHb) - 2.5) + 2.52\pm0.03$.

Galaxies with detected \lya tend to occupy the top right of the plot, with strong nebular emission lines, suggesting these are young star-forming galaxies with low metallicity and large ionization parameters, producing copious ionizing photons needed to power these emission lines \citep[see e.g.,][for more detailed studies]{Yang2017,Du2020,Tang2021}. We find the \lya selected galaxies have stronger \OiiiHb EW compared to the photometric population, following the trend with \Ha EW in Figure \ref{fig:xi_ion_Halpha}. Though, as discussed by \citet{Tang2021}, not all galaxies with strong nebular emission are detected in \lya, indicating that \lya transmission is reduced due to a high column density of neutral gas in these systems and/or inclination effects. We compare our data to a $z\sim2$ sample by \citet{Tang2019}, which was selected based on strong [OIII] emission. We find a similar correlation, but overall our ratio of \Ha EW/\OiiiHb EW is higher by $\sim$0.1 dex. Given that the \citet{Tang2019} sample has significantly sub-solar gas-phase metallicity $Z< 0.3Z_{\odot}$ \citep{Tang2021}, the decrease we observe in [OIII] at fixed \Ha EW would likely imply an overall lower metallicity due to a lower number of metal atoms in our sample.

\section{Discussion} \label{sec:disc}

\subsection{The profile of a strong ionizer}
\label{sec:disc_xiion}

Thanks to the depth of JWST/NIRCam we have been able to assess trends of \xiion at $z>3$ across the broadest range of galaxy properties to-date. From these results, we corroborate previous work at lower redshift and high luminosities, but push the measurement of \xiion to a  large sample of individual UV-faint galaxies for the first time. 

We find that galaxies with strong ionizing photon emission tend to have high \Ha EW, low UV luminosity, blue UV $\beta$ slope and \lya emission -- all implying that these galaxies are young, with likely low dust content and metallicity, and have a high O/B star population, capable of producing hard ionizing photons \citep[e.g.,][]{Tang2019,Boyett2022a}. This picture of the integrated emission from galaxies is complemented by high spatial resolution observations of highly magnified arcs with JWST. These have revealed extremely young star clusters ($\simlt10$\,Myr) with \OiiiHb EW $>1000$\,\AA, which dominate the ionizing photon production in their galaxy \citep{Vanzella2022a,Vanzella2022b}, indicating that there can be large variations in \xiion in individual galaxies, if they contain multiple stellar populations, but that the variation is primarily driven by the age of the stellar populations. We also find that overall, our \lya galaxy sample has higher \xiion than the photometrically selected one; the primary reason for this difference is that the former has higher \Ha EW (Figure~\ref{fig:xi_ion_Halpha}). The enhanced prevalence of \lya emission in strong \Ha emitters is likely a combination of increased production of \lya photons due to the young stellar population implied by strong \Ha, and potentially also an increase in the \lya escape fraction in the interstellar medium \citep[][]{Tang2021,Naidu2022a}. In these rapidly star forming galaxies, the hard ionizing radiation may be ionizing the ISM and/or feedback may disrupt the ISM gas leading to a reduced HI column density and dust cover. We note that the galaxies with the highest $(1-\fesc)\xiion$ are not necessarily all \lya-emitters, likely due to variance in the geometry and column density of neutral gas and dust in these sources. \citep{Ning2023} has shown this same correlation between \xiion and \lya for a broad range of luminosities and equivalent widths.

\subsection{The ionizing photon escape fraction}
\label{sec:disc_fesc}

In Section~\ref{sec:res_fesc_Z} we explored whether our sample shows signs of high ionizing photon escape fraction, \fesc, using the low H$\beta$ EW - blue UV $\beta$ slope region defined by \citet{Zackrisson2017} for $\fesc > 0.5$. While several sources fell into this region, with both low \Ha EW and \OiiiHb EW ($\simlt100$\,\AA), the uncertainties on the line flux measurements are too large for these to be robust candidates. More precise emission line measurements with JWST spectroscopy will be vital for identifying such candidates and their relative abundance in the galaxy population.

The lack of high \fesc candidates amongst the \lya-selected galaxies is also surprising. As the same conditions (low neutral gas covering fraction) facilitate both \lya escape and Lyman continuum escape, a correlation between \lya and Lyman continuum escape fraction is expected \citep[e.g.,][]{Verhamme2015,Dijkstra2016a,Reddy2016}. 

As discussed by \citet{Topping2022}, however, it is possible for galaxies with high \fesc but with very young ages to still have high nebular emission due to high ionizing photon production. It is likely that the criteria proposed by \citet{Zackrisson2017} can only find high \fesc systems within the bounds of the assumptions made for their model, such as galaxy SFH, ages, metallicities, dust, but also the stellar models used. Our results suggest the low luminosity galaxies with high sSFR but low \OiiiHb EW observed by \citet{Endsley2022} may be more likely due to variation in metallicity than high \fesc.

\section{Conclusions} \label{sec:conc}

We have inferred the hydrogen ionizing photon production rate, modulo the escape fraction, in the largest sample of individual sub-$L^*$ $z>3$ galaxies to-date, spanning $-23 \simlt \MUV \simlt -15.5$, with a median $\MUV=-18.1$, thanks to deep JWST/NIRCam imaging, enabling us to track the demographics of the ionizing population. Our conclusions are as follows:
\begin{enumerate}
    \item The median $\log_{10}(1-\fesc)\xiion$ of our sample is $25.33\pm 0.47$ with an intrinsic scatter of 0.42 dex. The inferred \xiion distribution of our sample has values in a range of $\sim1.5$\,dex, implying a wide range of galaxy properties and ages.
    \item We find significant trends of increasing $(1-\fesc)\xiion$ with increasing \Ha EW, decreasing UV luminosity, and with decreasing UV slope, all suggesting galaxies which are most efficient at producing ionizing photons are young, highly star forming, which are normally expected to be low metallicity and dust-poor.
    \item We find galaxies selected with strong \lya emission to have higher \xiion than photometrically-selected galaxies, with median $\log_{10}(1-\fesc)\xiion$ values of $25.39\pm0.64$ and $25.31\pm0.43$ respectively. We find the \lya-detected galaxies have an elevated \Ha EW distribution, thus the increased \xiion is likely driven by the selection based on \lya selecting a younger population. As strong \lya emitters also likely have high ionizing photon escape fractions, this implies the intrinsic production rate of ionizing photons in these galaxies could be significantly higher than what we can infer from \Ha luminosities.
    \item We examine our sample for signs of very high \fesc by comparing the inferred strengths of nebular emission lines (\OiiiHb and \Ha) and the strength of the nebular continuum via the UV $\beta$ slope. We find no significant evidence for sources with high escape fraction galaxies with low nebular emission line strength and very blue UV $\beta$ slopes. The reduced strength of \OiiiHb EW in our $z>3$ sample compared to a sample at $z\sim2$ from \citet{Tang2019} implies our sample has likely lower gas-phase metallicity and/or ionization parameter. 
\end{enumerate}

We have demonstrated the power of JWST/NIRCam photometry to more precisely constrain the rest-frame optical emission of UV-faint high redshift galaxies than previously possible with Spitzer/IRAC. These observations allow us to constrain the production rate of ionizing photons from early galaxies, corroborating the picture obtained from previous stacking analyses that \xiion is elevated in young, highly star forming galaxies, but that there is a broad distribution of \xiion, likely driven by variation in galaxy properties and ages. 

With JWST spectroscopy it is becoming possible to obtain direct measurements of optical emission lines in large samples \citep[e.g.,][]{Sun2022,Williams2022,Matthee2022a}. Deriving a census of the ionizing photon production rate across the full galaxy population will be necessary to fully understand reionization. Here we have shown that \xiion is elevated in UV-faint galaxies with strong nebular emission lines, likely due to young ages. While a thorough analysis of the implications of our results for reionization are beyond the scope of this work, this becomes more prominent at high redshift \citep[e.g.,][]{Boyett2022a,Endsley2022}, implying that it would be possible to complete reionization with modest \fesc. Considering the full distributions of \xiion and \fesc across galaxy properties will be required to assess the primary drivers of reionization.

\begin{acknowledgements}
We thank the co-PIs Ivo Labb\'{e} and Rachel Bezanson for the conception and public availability of the UNCOVER JWST Program (GO-2561), which made much of this work possible. We thank Mengtao Tang for sharing data the emission line catalog from \citet{Tang2019}. This work is based on observations collected at the European Southern Observatory under ESO programmes 109.24EZ.001 and 094.A-0115. This work is based on NASA/ESA HST and JWST data which were obtained from the Mikulski Archive for Space Telescopes at the Space Telescope Science Institute, which is operated by the Association of Universities for Research in Astronomy, Inc., under NASA contract NAS 5-03127 for JWST. The HST observations are associated with programs GO/DD-17231, GO/DD-13495, GO-15117, GO-11689, GO-11386, GO-13389, GO-15940 and SNAP-16729. The JWST observations are associated with programs JWST-ERS-1324 and GO-2561. CM and GP acknowledge support by the VILLUM FONDEN under grant 37459. The Cosmic Dawn Center (DAWN) is funded by the Danish National Research Foundation under grant DNRF140. We acknowledge financial support from NASA through grant JWST-ERS-1324. We acknowledge support from the INAF Large Grant 2022 “Extragalactic Surveys with JWST”  (PI Pentericci), and support through grants PRIN-MIUR 2017WSCC32, 2020SKSTHZ. This research is supported in part by the Australian Research Council Centre of Excellence for All Sky Astrophysics in 3 Dimensions (ASTRO 3D), through project number CE170100013.

\end{acknowledgements}

%
%

\bibliographystyle{aa}
\bibliography{library}

\end{document}